\documentclass[aps,prl,twocolumn,superscriptaddress,superscriptreference]{revtex4-2}
\usepackage{amsmath,bbold,bm,amssymb,scalerel,mathtools}
\usepackage{graphicx}
\usepackage{color}
\usepackage{enumitem}
\usepackage{algorithm,algpseudocode}
\usepackage{multirow}
\usepackage{colortbl,booktabs}
\usepackage{placeins}
\usepackage[usenames,dvipsnames]{xcolor}
\usepackage{tikz}

\usepackage{pdfpages}
\usepackage{pgffor}
\makeatletter
\AtBeginDocument{\let\LS@rot\@undefined}
\makeatother

\usepackage[colorlinks, linkcolor=blue!50!black, urlcolor=blue!50!black, citecolor=blue!50!black]{hyperref}
\tikzset{%
	every neuron/.style={
		circle,
		draw,
		minimum size=0.5cm
	},
	neuron missing/.style={
		draw=none, 
		scale=1.5,
		text height=0.333cm,
		execute at begin node=\color{black}$\vdots$
	},
}

\newcommand\equalhat{%
	\let\savearraystretch\arraystretch
	\renewcommand\arraystretch{0.3}
	\begin{array}{c}
		\stretchto{
			\scalerel*[\widthof{=}]{\wedge}
			{\rule{1ex}{3ex}}%
		}{0.5ex}\\ 
		=%
	\end{array}
	\let\arraystretch\savearraystretch
}

\newcommand{\gj}[1]{\textcolor{black}{#1}}

\usepackage[normalem]{ulem}

\begin{document}

\title{Predicting dynamic heterogeneity in glass-forming liquids by physics-\gj{inspired} machine learning}

\author{Gerhard Jung}

\affiliation{Laboratoire Charles Coulomb (L2C), Universit\'e de Montpellier, CNRS, 34095 Montpellier, France}

\author{Giulio Biroli}

\affiliation{Laboratoire de Physique de l’Ecole Normale Supérieure, ENS, Université PSL, CNRS, Sorbonne Université, Université de Paris, F-75005 Paris, France}

\author{Ludovic Berthier}

\affiliation{Laboratoire Charles Coulomb (L2C), Universit\'e de Montpellier, CNRS, 34095 Montpellier, France}

\affiliation{Yusuf Hamied Department of Chemistry, University of Cambridge, Lensfield Road, Cambridge CB2 1EW, United Kingdom}

\date{\today}

\begin{abstract}
We introduce GlassMLP, a machine learning framework using physics-\gj{inspired} structural input to predict the long-time dynamics in deeply supercooled liquids. We apply this deep neural network to atomistic models in 2D and 3D. Its performance is better than the state of the art while being more parsimonious in terms of training data and fitting parameters. GlassMLP quantitatively predicts four-point dynamic correlations and the geometry of dynamic heterogeneity. \gj{Transferability across system sizes allows us to efficiently probe} the temperature evolution of spatial dynamic correlations, revealing a profound change with temperature in the geometry of rearranging regions. 
\end{abstract}

\maketitle

Glasses are formed by the continuous solidification of supercooled liquids under cooling, while maintaining an amorphous microstructure~\cite{doi:10.1021/jp953538d}. Understanding glass formation and the phenomenon of the glass transition has been the focus of an intense research activity~\cite{berthier2016facets,CAVAGNA200951}. 

An important feature of supercooled liquids is the growth of spatial heterogeneity characterising the relaxation dynamics, where some regions actively rearrange while others appear completely frozen~\cite{bookDH}. Recently, an important effort was devoted to understanding the connection between dynamic heterogeneity and structural properties~\cite{Royall2015,tanaka2019revealing,PhysRevMaterials.4.113609}. Several structural parameters were shown to correlate with the dynamics, including density, potential energy~\cite{PhysRevLett.91.235501,Harrowell2006}, locally favored structures~\cite{Malins2013,tong2018revealing,hocky2014correlation}, but also more complicated quantities such as soft modes~\cite{isoconf2}, local yield stress~\cite{lerbinger2021relevance} and Franz-Parisi potential~\cite{PhysRevLett.127.088002}. The search intensified with the emergence of machine learning (ML) allowing the detection of correlations from unsupervised~\cite{unsupervisedFilion,unsupervisedCoslovich,CNN2022,PhysRevE.106.025308} or supervised~\cite{PhysRevLett.114.108001,schoenholz2016structural,bapst2020unveiling,Zaccone2021,Filion2021,Filion2022,GNNrelative2022} learning. The explored methodologies range from simple linear regression and support vector machines using a set of handcrafted structural descriptors~\cite{PhysRevLett.114.108001} to graph neural networks (GNN) with tens of thousands of adjustable parameters~\cite{bapst2020unveiling,GNNrelative2022}. Despite this versatility, \gj{none of the proposed networks can so far predict dynamic heterogeneities and related multi-point correlation functions that quantitatively agree with the actual dynamics.} 

Here, we \gj{bridge this major gap} by leveraging and combining previous ML approaches. We construct a physics-\gj{inspired} deep neural network that uses established structural order parameters as input to predict long-time dynamics in deeply supercooled liquids. The proposed methodology, which surpasses the state of the art, allows us to \gj{very efficiently} obtain quantitative predictions about heterogeneous dynamics and hence to gather \gj{novel} physical insights about their temperature evolution.

We simulate a Lennard-Jones non-additive mixture in 3D (KA, \cite{PhysRevE.51.4626}) for comparison with earlier work \cite{bapst2020unveiling} and a 2D ternary mixture (KA2D) \gj{where lower temperatures can be accessed.} We focus on KA2D since its interactions were adapted to efficiently prevent crystallization~\cite{Berthier2020} and enable the use of the swap Monte Carlo (SWAP) algorithm~\cite{swap:ninarello2017,swap:Berthier2019}. Equilibrium configurations are created with $N=1290$ particles ($M_\text{type}=3$, $N_1=600$, $N_2=330$, $N_3=360$) and box length $L=32.896$ using periodic boundary conditions and reduced units. We use SWAP to equilibrate the system and create a statistical ensemble. The average over equilibrium configurations is denoted $\langle \cdots \rangle$. For each configuration, $N_R=20$ replicas are created by drawing initial velocities from the Maxwell distribution to analyze the isoconfigurational ensemble~\cite{isoconf1,isoconf2} in which one averages over velocities at fixed initial configuration. We then simulate the dynamics using molecular dynamics (MD) and calculate for each particle $i$ the isoconfigurational average of the bond-breaking correlation function $\mathcal{C}^i_B(t) = \langle n^i_t/n^i_0 \rangle_\text{iso}$, which following \cite{isoconf1,isoconf2} we call ``propensity''; $\mathcal{C}^i_B(t)$ describes the number $n^i_t$ of nearest neighbors particle $i$ still has after a time $t$ relative to its $n^i_0$ initial number of neighbors~\cite{guiselin2022microscopic}. From the averaged propensity $\bar{\mathcal{C}}_B(t) = \frac{1}{N_1} \sum_{i \in N_1} \mathcal{C}^i_B(t)$, we extract a structural relaxation time, $\tau_\alpha^\text{BB}$, defined as $\langle \bar{\mathcal{C}}_B(t=\tau_\alpha^\text{BB}) \rangle = 0.5$. We report results for type 1 but verified that all findings are independent of particle type. We focus on three different temperatures: (i) slightly below the onset temperature ($T=0.4$, $\tau_\alpha^\text{BB}=1.7 \times 10^3$), (ii) slightly above the mode-coupling temperature ($T=0.3$, $\tau_\alpha^\text{BB}=3.4 \times 10^4$) and (iii) slightly below the mode-coupling temperature ($T=0.23$, $\tau_\alpha^\text{BB}=4.0 \times 10^6$). More details are given in  the Supplemental material (SM)~\cite{SM}.

\begin{figure}
	\hspace*{-0.2cm}\includegraphics[scale=0.99]{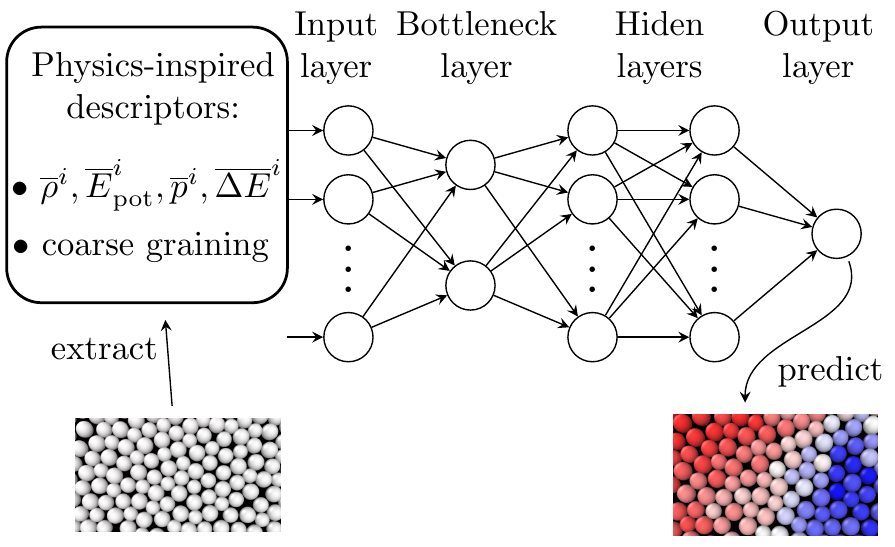}
	\caption{Sketch of the GlassMLP network. The physics-\gj{inspired} input is extracted from the initial inherent structure and inserted via the input layer. The network parameters are trained in a supervised learning procedure from propensities calculated using molecular dynamics simulations. After training, the network is able to predict the propensities of a new set of configurations (blue high propensity, red low one).}
	\label{fig:network}
\end{figure}

The first step in the ML approach is to select physics-\gj{inspired} inputs: \gj{a 
number $M_S$ of structural descriptors constructed for each particle $i$ from $K$ different observables. Inspired by the handcrafted features in Refs.~\cite{Filion2021,Filion2022} we also calculate coarse-grained averages of these descriptors on different length scales $L$. The first descriptor is the coarse-grained local density, $\overline{\rho}^i_{L,\beta} = \sum_{j\in N^i_\beta} e^{-R_{ij}/L}$, where the sum runs over all $N_\beta^i$ particles of type $\beta$ within distance $R_{ij} = |\bm{R}_i - \bm{R}_j| < 20$ of particle $i$. Particle positions are evaluated in the inherent structures $\bm{R}_i$. Similar in philosophy to Ref.~\cite{Zaccone2021} we additionally choose three different physics-\gj{inspired} descriptors: the coarse-grained potential energy,
$
   \overline{E}^i_{L,\beta} = \sum_{j\in N_\beta^i} E^j e^{-R_{ij}/L} / \bar{\rho}^i_{L,\beta} ,
$  extracted from the pair potential $E^i = \sum_{j\neq i} V(R_{ij})/2$,
the local Voronoi perimeter
$
   \overline{p}^i_{L,\beta} = \sum_{j\in N_\beta^i} p^j e^{-R_{ij}/L} / \overline{\rho}^i_{L,\beta} ,
$  using the perimeter $p^i$ of the Voronoi cell around particle $i,$ extracted using the software Voro++\cite{voro++}, and finally local variance of potential energy,
$\overline{\Delta E}^i_{L,\beta} = \sum_{j\in N_\beta^i} (E^j - \overline{E}_{L,\beta}^i)^2 e^{-R_{ij}/L} / \overline{\rho}^i_{L,\beta} $.} As coarse-graining lengths we choose $M_\text{CG}=16$ values $L=\{0.0,0.5,\dots,7.5\}$. In addition to coarse graining the descriptors separately for each of the $M_\text{type}$ types we also calculate the coarse-grained average by iterating over all particles independently of type. In total, this procedure therefore produces a set of $M_S = K M_\text{CG} (M_\text{type}+1) = 256$ descriptors. To simplify the learning, each descriptor is shifted and rescaled to have zero mean and unit variance over the training set.

We then apply a supervised ML procedure to train a multilayer perceptron (MLP) to give a prediction $\mathcal{X}_\text{MLP}^i$~\cite{haykin1994neural} for the propensity of particle $i$. Between the input and output layers, we introduce three hidden layers with 2, 10 and 10 nodes, respectively, as sketched in Fig.~\ref{fig:network}. In total, our model has around 650 fitting parameters, about 100 times less than the GNN proposed in Ref.~\cite{bapst2020unveiling}, and slightly fewer than the networks used in Refs.~\cite{Filion2021,Filion2022} due to a significant reduction in the number of structural descriptors $M_S$. The intermediate layer with only 2 nodes is a bottleneck layer. Its introduction is crucial to prevent overfitting of the training data and represents a major difference to the MLP suggested in Ref.~\cite{Filion2022} where unsatisfying results were reported. We name our deep neural network `GlassMLP'. We use $N_S=300$ initial structures, which are equally divided into training, validation and test sets. During learning, we compute for each configuration as loss function the mean absolute error between true and predicted labels~\cite{bapst2020unveiling,Filion2021,Filion2022}. In the loss we also include terms that penalize deviations from the true variance and spatial correlations of the propensities. Both quantities are evaluated by averaging over all particles in the configuration for which the loss function is evaluated. For the training we apply stochastic gradient descent with an Adam optimizer~\cite{Adam}. The hyperparameters used for training are the same for all times and temperatures. \gj{The training of GlassMLP on one state point requires less than five minutes on a Laptop GPU (NVIDIA T600 Laptop).}

\begin{figure}	\hspace*{-0.25cm} \includegraphics{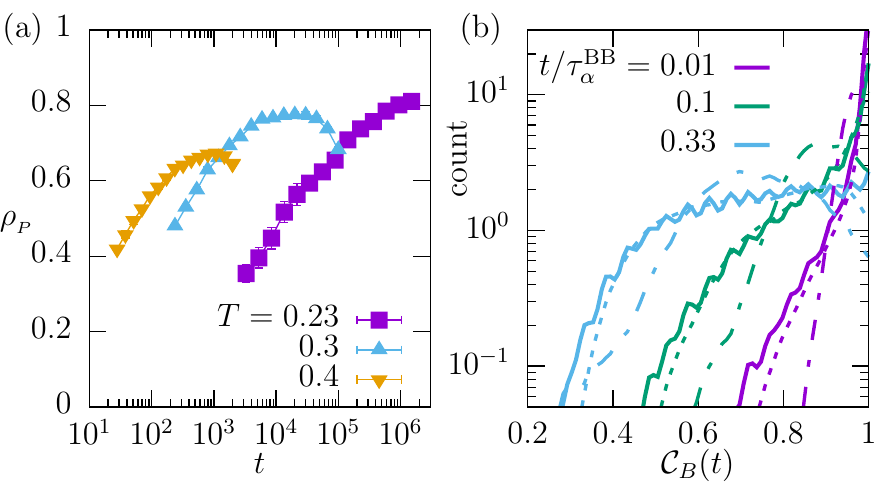}
	\caption{Performance of GlassMLP applied to the KA2D model. (a) Time evolution of the Pearson correlation between GlassMLP predictions and MD results for different temperatures. (b) Probability distributions of propensity calculated from MD (full line), GlassMLP (dotted line), and Ridge regression (dashed-dotted line) for different time scales at $T=0.23$. }
	\label{fig:pearson_histogram}
\end{figure}

To quantify the performance of GlassMLP we compute the Pearson correlation coefficient $\rho_P = \text{cov}(\mathcal{C}^i_B, \mathcal{X}^i_\text{MLP}) / \sqrt{\text{var}(\mathcal{C}^i_B) \text{var} (\mathcal{X}^i_\text{MLP})}$, between the true propensities $\mathcal{C}^i_B$ and the network output, $\mathcal{X}^i_\text{MLP}$. Perfect predictions would yield $\rho_P=1$ while random ones correspond to $\rho_P=0$. As shown in Fig.~\ref{fig:pearson_histogram}a, we find that $\rho_P$ depends non-monotonically on time and is maximal around $t \approx \tau_\alpha^\text{BB}/3$. Furthermore, the predictability considerably increases at lower temperatures and reaches values up to $\rho_P > 0.8$, which is significantly better than previously proposed techniques on KA models~\cite{unsupervisedFilion,unsupervisedCoslovich,bapst2020unveiling,Filion2021,Filion2022}. \gj{A direct comparison to GNNs \cite{bapst2020unveiling} is presented below for the 3D KA model.} 

\gj{We now go beyond establishing the quality of a correlation and focus on the probability distribution of the propensity.} Fig.~\ref{fig:pearson_histogram}b shows an excellent agreement between GlassMLP predictions and  MD results. Minor discrepancies exist in the tails for small propensities, as the network slightly underestimates variances. Poor results are instead obtained by   the Ridge regression method suggested in \cite{Filion2021,Filion2022}, which  always outputs nearly Gaussian distributions. This shows that using a \gj{non-linear} neural network such as GlassMLP is important to capture the complex shape of the distributions. \gj{See SM for further comparison between methods \cite{SM}.} 

\begin{figure}
\hspace*{-0.2cm}	\includegraphics{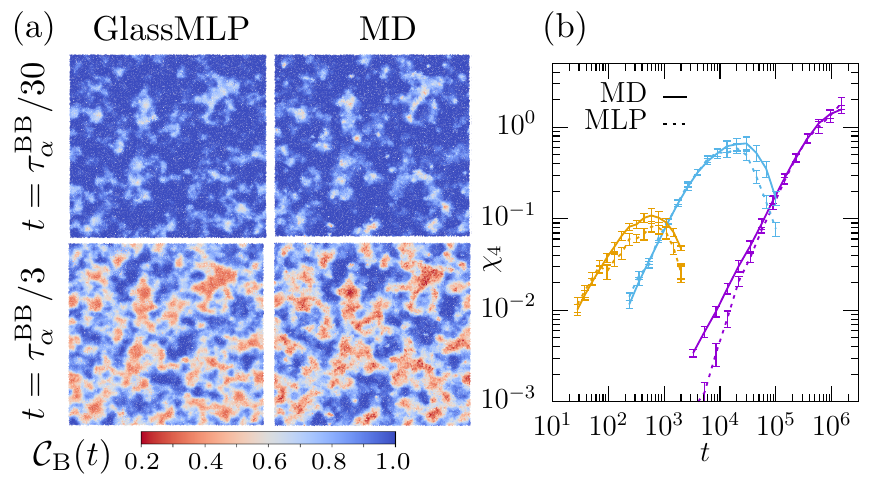}
		\caption{Dynamic heterogeneities in MD simulations and GlassMLP. (a) Snapshots of an representative configuration for different time scales at $T=0.23$, where blue regions with high propensity move very little. (b) Susceptibility ${\chi}_4(t)$ against time $t$ for different temperatures as in Fig~\ref{fig:pearson_histogram}. Further snapshots in SM. }
	\label{fig:snapshots_chi4}
\end{figure}

\gj{Because GlassMLP performs excellently at the single-particle level, we now apply it to spatial correlations, thus promoting GlassMLP as a new tool to probe dynamic heterogeneity~\cite{PhysRevE.76.041509}.} First, we show snapshots of the predicted and calculated propensities for different time scales in Fig.~\ref{fig:snapshots_chi4}a. The MD results show how marginally rearranged active clusters at small times (white and red) coarsen with time and become both larger and more strongly contrasted to the unrelaxed background (blue)~\cite{Berthier2022}.  GlassMLP is able to predict remarkably well the location and the geometry of the relaxing clusters from the sole knowledge of the initial structure.

Spatially heterogeneous dynamics is quantified by the four-point susceptibility $\chi_4(t)= N_1 \left( \langle \bar{\mathcal{C}}_B^2(t)  \rangle - \langle \bar{\mathcal{C}}_B(t)  \rangle^2  \right)$  shown in Fig.~\ref{fig:snapshots_chi4}b. Its time dependence is similar to the one of the Pearson correlation, with a maximum at $t \approx \tau_\alpha^\text{BB}/3$ that grows upon cooling. This similarity suggests that GlassMLP is particularly powerful in analysing strongly heterogeneous dynamics. \gj{The effect is further enhanced due to the increased structural origin for dynamic heterogeneities at lower temperatures observed in earlier work \cite{PhysRevE.76.041509}. Fig.~\ref{fig:snapshots_chi4}b also highlights} that GlassMLP accurately predicts the time and temperature evolution of $\chi_4(t)$. To our knowledge, no ML technique has previously been able to predict $\chi_4(t)$ at a comparable quantitative level. This susceptibility quantifies the average number of correlated particles during structural relaxation~\cite{PhysRevE.71.041505} and can be accessed experimentally~\cite{berthier2005direct,PhysRevE.76.041510}.

\begin{figure}
	\includegraphics{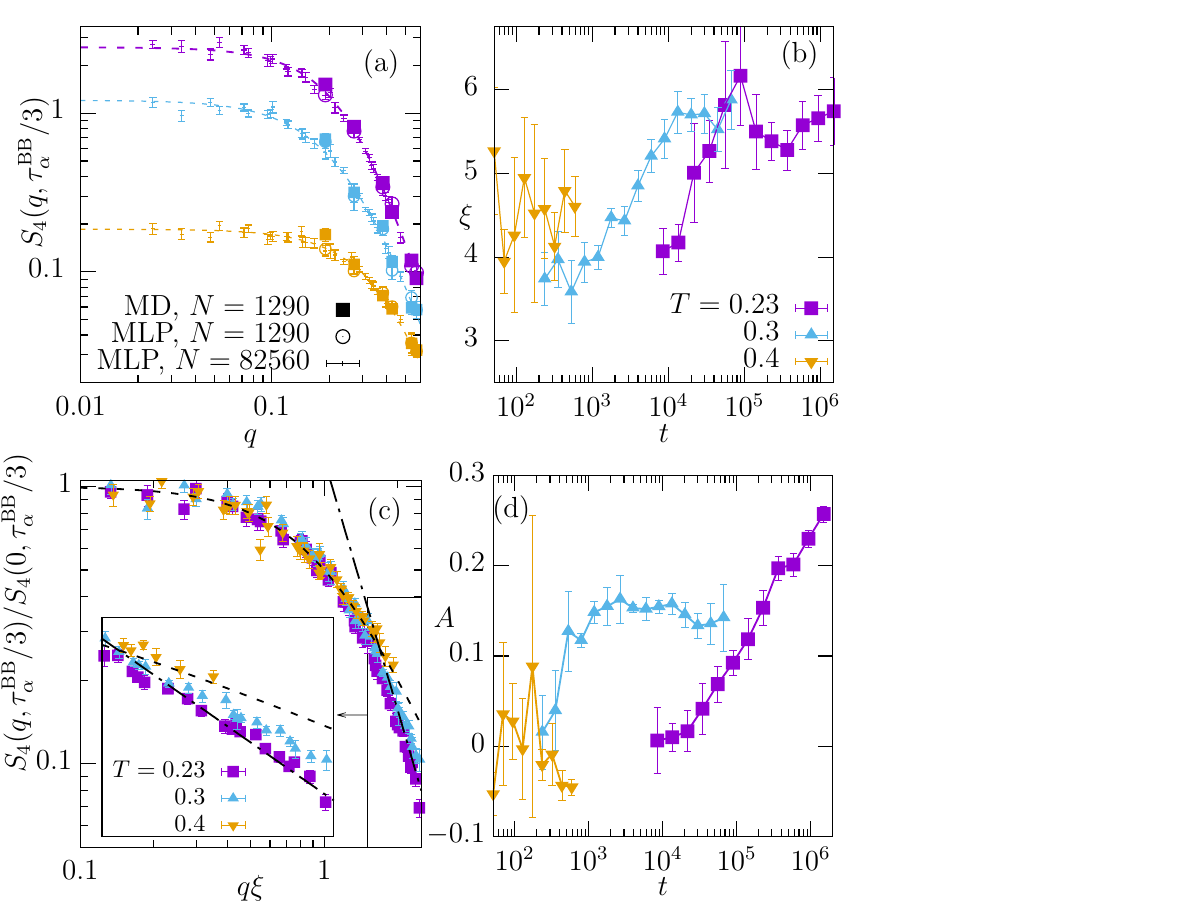}
	\caption{Evolution of length scales and geometry of dynamic heterogeneity in the 2DKA model. (a) Four-point structure factor slightly below the structural relaxation time $\tau_\alpha^\text{BB}/3$ for different temperatures $T$ and system sizes $N$. \gj{Dashed lines are fits $S_4(q,t) = \Tilde{\chi}_4(t) / \left( 1 +  (\xi q)^2 + A(\xi q)^3\right )$ as rationalized in the main text.} (b) Length scales $\xi$ extracted from non-linear fits described in the main text. Only points for which the Pearson coefficient $\rho_{_P} > 0.5$ are shown. (c) Rescaled four-point structure factor vs rescaled wavenumber $q\xi$ for the MLP, $N=82560$ data. Dashed lines corresponds to $(1+(q\xi)^2)^{-1}$ and dashed-dotted line is $\sim q^{-3}$. Inset shows zoomed data for large $q\xi$. (d) Higher-order prefactor A, extracted from fitting $S_4(q,t)$ as described in the main text.}
	\label{fig:S4}
\end{figure}

The evolution of $\chi_4(t)$ results from two factors~\cite{Berthier2007_1,Berthier2007_2}: a growing length scale characterising the decay of dynamic correlations, and a growing strength of these correlations. \gj{We now show that GlassMLP can even disentangle them.} Let us define the four-point structure factor, $S_4(q,t) = N_1^{-1} \left\langle W(\bm{q},t) W(-\bm{q},t) \right\rangle$, with $W(\bm{q},t) = \sum_{i \in N_1}  (\mathcal{C}_B^i(t) - \langle \bar{\mathcal{C}}_B(t) \rangle) \exp[\textrm{i} \bm{q}\cdot \bm{R}_i(0)]$. See SM for the analysis of its real space counterpart. The measured $S_4(q,t)$, shown in Fig.~\ref{fig:S4}a, displays a peak at small $q$ which contains all relevant information about spatial dynamic correlations. For this function the predictions made by GlassMLP are again in excellent agreement with measurements.
It is notoriously difficult to quantitatively extract a correlation length scale $\xi$ from $S_4(q,t)$ as one needs systems much larger than $\xi$~\cite{karmakar2010analysis,Flenner2010,karmakar2014growing,Flenner2016}. Previous works tackled this challenge by simulating very large systems which becomes a real challenge at low temperatures where long time scales are also needed. GlassMLP fully solves this problem  by transferring results from small to large systems. One can train GlassMLP on reasonably small \gj{(but not too small)} systems and then apply it to  very large ($N=82560$) equilibrium configurations obtained using SWAP. GlassMLP predicts the propensity field and hence $S_4(q,t)$ for these configurations at essentially no cost because the network is already trained and the slow dynamics of large systems is never simulated. \gj{The transferability in system size is possible because the bond-breaking correlation and $S_4(q,t)$ has been shown to be independent of system size for the chosen $N$ values~\cite{PhysRevLett.117.245701,Flenner2016}. See SM for finite-size analysis~\cite{SM}.} This method allows us to obtain for the first time reliable data for $S_4(q,t)$ over an extended range of times, temperatures, and wave vectors, see Fig.~\ref{fig:S4} \footnote{ \gj{The data analyzed corresponds roughly to an equivalent of $5 \cdot 10^6$ CPU hours when using standard MD simulations.}}. We find that an Ornstein-Zernicke functional form, $S_4 \approx 1/(1+(q\xi)^2)$ does not describe the numerical data over the entire range of temperature \gj{for $q>0.2$} and a higher-order term is needed. This was proposed theoretically using mode-coupling theory~\cite{PhysRevLett.97.195701} with a quartic term, and in the East model~\cite{3DEast} where a fractal exponent $q^{0.58+D}$ is found. Neither proposal is consistent with our data. Because dynamic heterogeneity appears increasingly contrasted with more compact boundaries at lower temperatures~\cite{Berthier2022}, we introduce a cubic term $q^3$ by analogy with Porod's law describing two-phase systems with sharp interfaces~\cite{bray2002theory}: $S_4(\gj{0.2<q<0.6},t) = \Tilde{\chi}_4(t) / \left( 1 +  (\xi q)^2 + A(\xi q)^3\right )$. This expression contains the minimal ingredients to describe both the evolution of the characteristic length scale $\xi$ (Fig.~\ref{fig:S4}b) and of the geometry of dynamic heterogeneity (Figs.~\ref{fig:S4}c,d). The correlation length shows a maximum slaved to $\tau^{\rm BB}_\alpha$, which grows as temperature decreases. The temperature dependence is relatively weak, which stems from both the use of the bond-breaking correlation~\cite{Flenner2016} and of the isoconfigurational average~\cite{PhysRevE.76.041509,Poole2010,dunleavy2015mutual}. Interestingly, the prefactor $A$ is essentially zero at high temperature, but grows to dominate the $q$-dependence of $S_4$ at low $T$. These results reveal that at lower temperatures interfaces separating dynamically \gj{correlated domains become sharper while the domains become geometrically more compact~\cite{stevenson2006shapes,DAS2022100098,Berthier2022}.}

\begin{figure}	\hspace{-0.15cm}\includegraphics[]{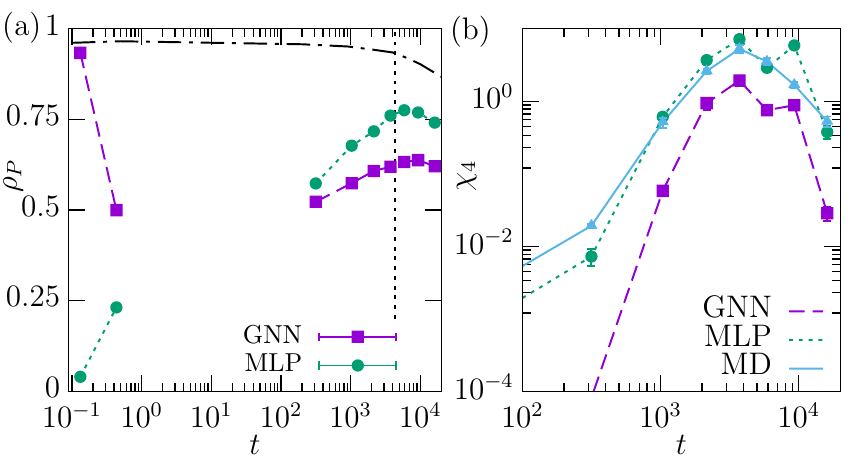}
	\caption{Comparison of two different ML techniques to predict the isoconfigurational average of displacements, $\mathcal{R}(t)$, for the 3D KA model. (a) Pearson correlation coefficient $\rho_P$ for different times $t$ at temperature $T=0.44$. The vertical dotted line marks structural relaxation $t=\tau_\alpha$ and the dashed-dotted line is the maximal achievable correlation. (b) Susceptibility $\chi_4(t)$ compared to the ground truth (MD).}
	\label{fig:KA}
\end{figure}

We close with a brief analysis of the 3D KA model to which the GNN of Ref.~\cite{bapst2020unveiling} was initially applied. 
The aim is to compare GlassMLP and the GNN and to show the performance of GlassMLP for a different model. \gj{For the benchmarking, and to present a fair comparison, we use the same dataset and the pretrained GNNs provided by Ref.~\cite{bapst2020unveiling} and similarly define the propensity as the isoconfigurational average of particle displacements, $\mathcal{R}^i(t)$, instead of $\mathcal{C}^i_B(t)$}. The setup for GlassMLP is as in 2D, we simply replace the perimeter $p^i$ with the surface area $s^i$ from the Voronoi decomposition.
Comparing the performance of GlassMLP with the GNN at $T=0.44$  in Fig.~\ref{fig:KA}a using the Pearson correlation coefficient $\rho_P$, we confirm that our network performs much better near structural relaxation while having less fitting parameters (factor of 100) and requiring less training data (factor of 10). Importantly, the improvement in performance is more obvious in the susceptibility $\chi_4(t)$ in Fig.~\ref{fig:KA}b which shows much better agreement with the MD result than the GNN, confirming GlassMLP as a versatile tool to analyse dynamic heterogeneity in glass-formers.  Very recent work~\cite{GNNrelative2022} on GNNs using relative particle motion and learning on edges instead of vertices was shown to yield Pearson correlations at the structural relaxation time comparable to ours, but no information was provided regarding dynamic heterogeneity.

In summary, we have developed GlassMLP, a deep neural network which uses physics-\gj{inspired} descriptors as input to predict long-time structural relaxation solely from the initial  structure. Improved performance is reached from (i) using prior knowledge about glass transition physics as inductive bias for neural networks~\cite{Zaccone2021}; (ii) including spatial correlations into the loss function; (iii) adjusting the architecture of the deep neural network to avoid overfitting.
Using transferability across system sizes allows to extract physically meaningful four-point dynamical structure factors and to analyse their physical evolution when approaching the glass transition. \gj{Although GlassMLPs performance is remarkable, the trained networks do not detect any outstanding features, which is consistent with the conclusions in Ref.~\cite{doi:10.1063/5.0128265}.} The success of GlassMLP \gj{therefore} demonstrates the importance of combining physics-\gj{inspired} inputs and deep neural networks able to extract \gj{inherent complex and non-linear features from them, with relative weights that are presumably model-dependent.}  

The method proposed here could easily be extended to include further descriptors and \gj{applied} to other types of systems, including experiments on glass-forming colloidal liquids \gj{or granular glasses}, where potentially different descriptors can be used. Our findings on spatially-correlated dynamics pave the way for more rigorous analysis of dynamic heterogeneity in deeply supercooled liquids to better understand their physical origin, and the interplay between heterogeneous structure~\cite{PhysRevLett.127.088002} and dynamic facilitation~\cite{Berthier2022} close to the experimental glass transition.

\acknowledgments  

We thank V. Bapst for providing simulation data and learned GNN models from Ref.~\cite{bapst2020unveiling} and C. Scalliet for guidance with the swap Monte Carlo LAMMPS code and explanations. We thank A. Liu, R. Chacko and S. Ridout for discussions. This work is supported by the Simons Foundation (\#454933, LB, \#454935 GB) and by a Visiting Professorship from the Leverhulme Trust (VP1-2019-029, LB).

\bibliography{library_local.bib}

\newpage

\newpage



\section*{Supplementary material}

# Supplemental Material "Predicting dynamic heterogeneity in glass-forming liquids by physics-inspired machine learning"

Gerhard Jung,[1] Giulio Biroli,[2] and Ludovic Berthier[1,3]

[1]*Laboratoire Charles Coulomb (L2C), Université de Montpellier, CNRS, 34095 Montpellier, France*
[2]*Laboratoire de Physique de l'Ecole Normale Supérieure, ENS, Université PSL,
CNRS, Sorbonne Université, Université de Paris, F-75005 Paris, France*
[3]*Yusuf Hamied Department of Chemistry, University of Cambridge,
Lensfield Road, Cambridge CB2 1EW, United Kingdom*

## I. COMPUTER SIMULATIONS: KA2D MODEL

We simulate a modified Kob-Andersen mixture in 2D (KA2D) interacting via a Lennard-Jones potential,

$$V_{\alpha\beta}(R_{ij}) = \begin{cases} 4\epsilon_{\alpha\beta}\left[\left(\frac{\sigma_{\alpha\beta}}{R_{ij}}\right)^{12} - \left(\frac{\sigma_{\alpha\beta}}{R_{ij}}\right)^6 + C_0 \right. \\ \left. + C_2\left(\frac{r_{ij}}{\sigma_{\alpha\beta}}\right)^2 + C_4\left(\frac{R_{ij}}{\sigma_{\alpha\beta}}\right)^4\right] & R_{ij} < R_{\alpha\beta}^{\text{cut}} \\ 0 & \text{otherwise.} \end{cases}$$

The KA2D system has been specifically developed for this manuscript and is closely related to the family of KA2 models suggested in Ref. [1]. The KA2D model is a ternary mixture $\alpha, \beta = \{1, 2, 3\}$ where types 1 and 2 interact via the usual Kob-Andersen non-additive interactions, $\epsilon_{11} = 1.0$, $\epsilon_{12} = 1.5$, $\epsilon_{22} = 0.5$ and $\sigma_{11} = 1.0$, $\sigma_{12} = 0.8$, $\sigma_{22} = 0.88$. Compared to the KA model, however, we introduce an intermediate third species with interaction, $\epsilon_{13} = 0.75$, $\epsilon_{23} = 1.5$, $\epsilon_{33} = 0.75$ and $\sigma_{13} = 0.9$, $\sigma_{23} = 0.8$, $\sigma_{33} = 0.94$. The cutoff is species-dependent $R_{\alpha\beta}^{\text{cut}} = 2.5\sigma_{\alpha\beta}$ and we set $C_0 = 0.04049023795$, $C_2 = -0.00970155098$ and $C_4 = 0.00062012616$ to make the potential continuous up to the second derivative. The above energy and length scales were empirically adjusted to minimize any signatures of local crystallization and lead to well-mixed, disordered structures while preserving the efficiency of the swap Monte Carlo algorithm. Results are reported in reduced Lennard-Jones units defined by $\epsilon_{11}$ (energy scale), $\sigma_{11}$ (length scale) and $\sigma_{11}\sqrt{m/\epsilon_{11}}$ (time scale), with mass $m = 1$ for all types.

### A. Equilibration and simulation

The advantage of the KA2D is that the swap Monte Carlo (SWAP) algorithm can be used in combination with molecular dynamics (MD) simulations, which significantly reduces relaxation times and therefore allows to create equilibrium configurations at very low temperatures [2, 3]. Configurations are created with $N = 1290$ particles ($N_1 = 600$, $N_2 = 330$, $N_3 = 360$) and box length $L = 32.896$. We equilibrate the model using SWAP until it reaches a steady state, making sure that the time-averaged self-intermediate scattering function of the SWAP dynamics

$$\varphi_s(t) = \left\langle N_\beta^{-1} \sum_{i \in N_\beta} \exp^{-i\mathbf{q}\cdot(\mathbf{R}_i(t+T) - \mathbf{R}_i(T))} \right\rangle, \quad (1)$$

does not evolve anymore with $T$. We choose $\mathbf{q} = (64\pi/L, 0)$ and calculate the structural relaxation time, $\tau_\alpha^{\text{SWAP}}$, defined as $\varphi_s(\tau_\alpha^{\text{SWAP}}) = e^{-1}$. After equilibration, we extract independent configurations every $t = 5\tau_\alpha^{\text{SWAP}}$ timesteps.

During equilibration and production, SWAP phases are included every 10 MD steps. Each SWAP phase consists of 2100 SWAP moves, in which a randomly selected particle is attempted to be swapped with another randomly selected particle of different type and accepted with a Metropolis criterion [3]. The MD stepsize is 0.01 and we employ a Nosé–Hoover thermostat [4]. After production, the SWAP algorithm is turned off to simulate physical relaxation dynamics for each initial condition. All simulations are performed using Lammps [5].

### B. Inherent structures

We empirically found that removing thermal fluctuations from the structural input significantly improves the performance of the ML model. Every initial configuration $\{\mathbf{R}_i(0)\}$ is quenched to its inherent structure before the structural observables for the machine-learning input are calculated. The inherent structure is generally identified as the nearest local energy minimum from a given thermal configuration. Here, we employ a simple steepest decent along the force gradient with very small stepsize $\Delta R = 0.001$. This ensures that particles do not move significantly during minimization. We perform minimization up to machine precision.

### C. Dynamics and time scales

We perform molecular dynamics simulations starting from each configuration and extract the bond breaking correlation coefficient $C_B^i(t) = n_t^i / n_0^i$ [6]. Here, $n_0^i$ is the number of particles $j$ within a cutoff $r_{\text{cut}}^0 = 1.4\sigma_{\alpha_i\beta_j}$ of particle $i$ and $n_t^i$ the number of particles that were initially part of the $n_0^i$ neighbors and are still inside a cutoff

$r_{\text{cut}}^t = 1.8\sigma_{\alpha_i\beta_j}$ at time $t$. The larger cutoff is chosen to ensure that particles really leave their cage and not just slightly fluctuate. To characterize the dynamics, we calculate the time scale, $\tau_\alpha^{\text{BB}}$, at which particles lose on average half of their neighbors, $\frac{1}{N_1}\left\langle \sum_{i\in N_1} C_B^i(t=\tau_\alpha^{\text{BB}}) \right\rangle = 0.5$. We also report the usual structural relaxation time $\varphi_s(\tau_\alpha^{\text{ISF}}) = e^{-1}$ of the MD simulations from the intermediate scattering function and evaluate the sixfold bond-orientational order parameter,

$$\Psi_i(t) = \frac{1}{N_n^i} \sum_{j=1}^{N_n^i} e^{\mathrm{i}6\theta_{ij}(t)}. \qquad (2)$$

Here, the sum runs over all neighbors $j$ of particle $i$ with distance $R_{ij}(t) = |\mathbf{R}_i(t) - \mathbf{R}_j(t)| < 1.4$. The number of these neighbors is denoted as $N_n^i$ and $\theta_{ij}(t)$ is the angle between the $x$-axis and $\mathbf{R}_i(t) - \mathbf{R}_j(t)$. The bond-orientational correlation function is then calculated as,

$$C_\Psi(t) = \left\langle \frac{\sum_{i\in N_\beta} \Psi_i(t)\Psi_i(0)^*}{\sum_{i\in N_\beta} |\Psi_i(0)|^2} \right\rangle, \qquad (3)$$

and used to extract the relaxation time, $C_\Psi(\tau_\alpha^{\text{BO}}) = e^{-1}$.

The different time scales are compared in Fig. 1. It can be seen that SWAP indeed leads to a significant speed up of more than 4 orders of magnitude at the lowest investigated temperature. Moreover, we find that $\tau_\alpha^{\text{ISF}}$ and $\tau_\alpha^{\text{BO}}$ are nearly identical. The absence of significant Mermin-Wagner fluctuations [7] is likely caused by the small system size. The bond-breaking time scales are always larger than the others, as discussed before Ref. [6]. The longest time scale reported for $T = 0.23$ in the main manuscript therefore corresponds to $\approx 2\tau_\alpha^{\text{ISF}}$. We do not find any strong dependence of these time scales on particle type, see Fig. 1.

From these characteristic time scales, we follow earlier reasoning [2] and provide estimates of the following characteristic temperature scales:

- onset temperature $T_o \approx 0.5$,
- mode-coupling temperature $T_{\text{MCT}} \approx 0.3$
- glass transition temperature $T_g \approx 0.15$.

## II. MACHINE-LEARNING MODEL

In the main manuscript we present the network structure of GlassMLP. Here, we will give some complementing details.

### A. Activation function

Each node of each layer is connected with all nodes of the adjacent layer. The value on each node $n$ in layer $l$

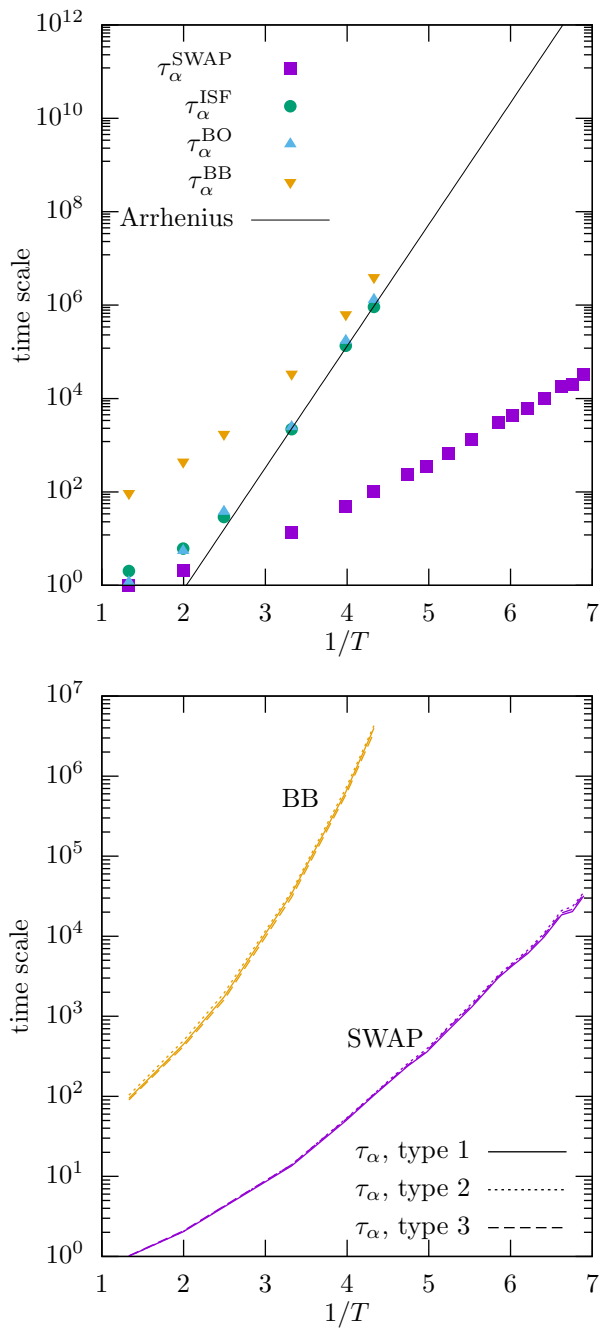

FIG. 1. Relaxation time scales extracted for various different observables from the SWAP simulations and the MD dynamics. Top: Time scales for type 1, including an Arrhenius fit for the three lowest temperatures to approximate $T_g$. Bottom: Time scales for the three different types of the KA2D model.

is calculated as

$$X_n^{(l)} = F^{\text{act}}\left( \sum_{m\in M^{(l-1)}} w_{mn}^{(l)} X_m^{(l-1)} + b_n^{(l)} \right), \qquad (4)$$

with nodes $M^{(l)}$ in layer $l$, learnable weights $w_{mn}^{(l)}$ and biases $b_n^{(l)}$ as well as non-linear activation function $F^{\text{act}}(X)$. For the latter we choose the exponential-linear unit (ELU) [8],

$$F_{\text{ELU}}^{\text{act}}(X) = \begin{cases} X & X \geq 0 \\ e^X - 1 & X < 0. \end{cases} \quad (5)$$

Only for the last layer we use a linear activation $F_{\text{lin}}^{\text{act}}(X) = X$ which reduces the values of the last hidden layer to one output value, i.e. the predicted propensity for particle $i$.

### B. Loss function

To train the network and find suitable weights and biases, the output of the network is rated by a loss function $L(\{\mathcal{X}_{\text{MLP}}\}, \{\mathcal{X}_{\text{MD}}\})$, where $\{\mathcal{X}_{\text{MD}}\}$ denotes the set of input labels obtained from MD simulations and $\{\mathcal{X}_{\text{MLP}}\}$ the set of network outputs in one batch. The loss is defined as

$$\begin{aligned} L = & N_{\text{batch}}^{-1} \sum_{i \in N_{\text{batch}}} |\mathcal{X}_{\text{MLP}}^i - \mathcal{X}_{\text{MD}}^i| / \sqrt{\text{Var}(\{\mathcal{X}_{\text{MD}}\})} \\ & + w_v \left( \text{Var}(\{\mathcal{X}_{\text{MLP}}\}) - \text{Var}(\{\mathcal{X}_{\text{MD}}\}) \right) / \text{Var}(\{\mathcal{X}_{\text{MD}}\}) \\ & + \sum_d w_d |C(\{\mathcal{X}_{\text{MLP}}\}, d) - C(\{\mathcal{X}_{\text{MD}}\}, d)| / C_N. \end{aligned} \quad (6)$$

The first line of this equation corresponds to a standard mean absolute error evaluated over all particles in a batch. The second line applies additional loss to deviations between the true and predicted propensity variances and the third line introduces a contribution to the loss to rate spatial correlations $C(\{\mathcal{X}\}, d)$ of the propensity. These are defined as,

$$C(\{\mathcal{X}\}, d) = \frac{\sum_{i,j=1}^{N_{\text{batch}}} \delta \mathcal{X}^i \delta \mathcal{X}^j e^{(R_{ij}-d)^2/2}}{\sum_{i=1}^{N_{\text{batch}}} (\delta \mathcal{X}^i)^2}, \quad (7)$$

where $\delta \mathcal{X}^i = \mathcal{X}^i - N_{\text{batch}}^{-1} \sum_i \mathcal{X}^i$. Both $C(\{\mathcal{X}\}, d)$ and $\text{Var}(\{\mathcal{X}\})$ are computed by averages over the particles in the configuration for which the loss function is evaluated. The Gaussian function in Eq. (7) is necessary to enable differentiation of the loss function and thus learning of the network via backpropagation. All three contributions to the loss function are normalized to give roughly equal contributions across temperatures and time scales. This allows us to choose the same hyperparameters for all systems and state points. The normalization for the spatial correlation is chosen as $C_N = |C(\{\mathcal{X}_{\text{MD}}\}, d_{\min}) - C(\{\mathcal{X}_{\text{MD}}\}, d_{\max})|$ and $d = 2, 4, 6$.

To train the network, we calculate propensity for $N_S = 300$ different initial structures, which are equally divided into a training, a validation and a test set. The batch size is set equal to the number of type 1 particles per configuration, $N_{\text{batch}} = N_1$. For the training we use an Adam optimizer [9]. The training is separated into different phases. In the first phase, the model is trained for 300 epochs with weights $w_d = w_v = 0.0$ and an accuracy of the Adam optimizer of $5 \times 10^{-4}$. Then we train with the same weights for 1000 epochs but include an early stopping using the validation loss with a patience of 15 epochs. The accuracy of the Adam optimizer is $2 \times 10^{-4}$. The results of this intermediate network are used in the comparison in Sec. III. Afterwards, the accuracy is further reduced to $4 \times 10^{-5}$ and the weights are set to $w_d = 0.5$ for all $d$ and $w_v = 1.0$. The network is trained for 50 more epochs. In the last stage it is again trained for another 1000 epochs using early stopping, as described above, and accuracy $2 \times 10^{-5}$.

## III. COMPARISON: GLASSMLP, MLP (MAE) AND RIDGE REGRESSION

In the main manuscript we present two advancements compared to the state-of-the-art ML algorithms: (i) we have introduced physics-inspired descriptors and (ii) we have used a more complex network structure and loss function. Here, we analyze the impact of both extensions. We present results using the same network structure as in the main manuscript (MLP with bottleneck) but a simplified loss function including only the mean-averaged error (mae). We will call this approach MLP (mae). We also fit the physics-inspired descriptors to the propensity using a simple Ridge regression [10].

We find that the results are very similar on the level of the Pearson correlation $\rho_P$ (see Fig. 2a). MLP (mae) is systematically slightly better than GlassMLP, because it has a more specified loss function to optimize the correlation coefficient. Ridge regression is slightly worse, but still much better than the GNN [11] or Ridge regression using different descriptors [10]. This is different for the susceptibility $\chi_4$ where GlassMLP is closer to the underlying MD results than the other two approaches (see Fig. 2b). When analyzing the four-point structure factor $S_4(q, t)$ we observe that GlassMLP clearly outperforms MLP (mae) and Ridge, in particular for smaller times. There is a small shift between GlassMLP and MD, which is due to the slight underestimation of the strength of the propensity fluctuations, as discussed in the main text. Apart from this shift, GlassMLP and MD show the same $q$-dependence. This is clearly not the case for MLP (mae) and Ridge which both decay much stronger, i.e. they predict larger length scales. Additionally, we have already shown that the distribution of propensities can not be properly predicted by the Ridge regression and remains Gaussian (see Fig. 2b in the main manuscript).

We conclude that the introduction of the physics-inspired descriptors are essential to improve the predictive power of the ML methodology on the level of the Pearson correlation coefficient $\rho_P$. To achieve quantitative predictions for propensity distributions, dynamic heterogeneities and length scales, it is additionally im-

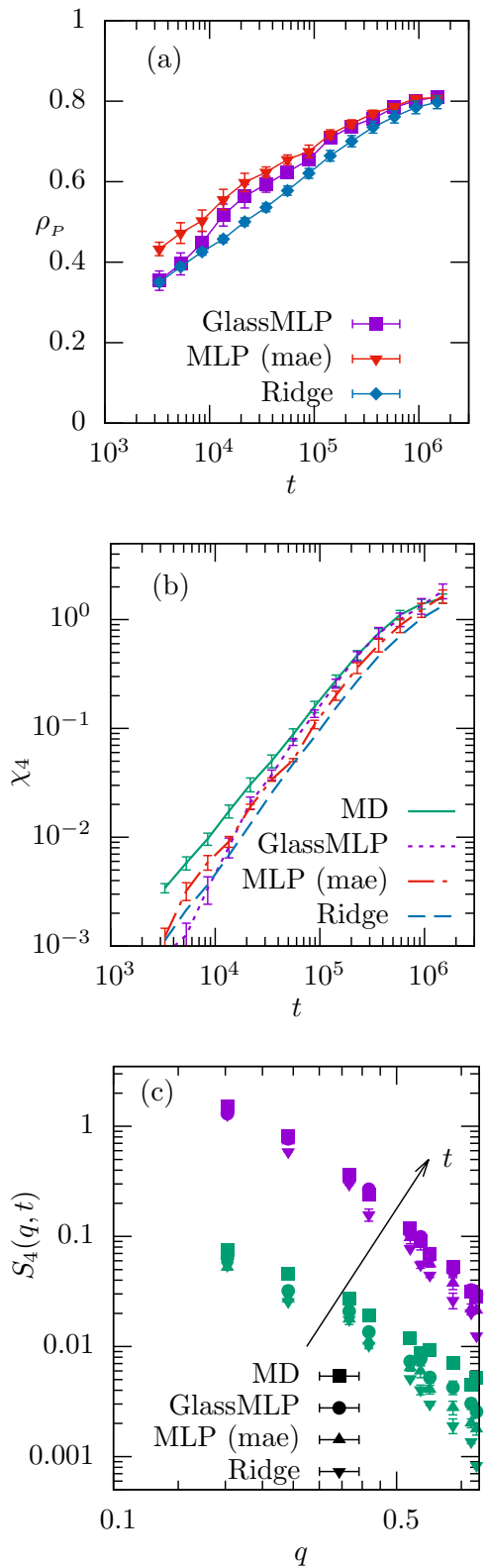

FIG. 2. Comparison between GlassMLP, as analyzed in the main manuscript and two other ML methodologies at $T = 0.23$. MLP (mae) only considers the mean-averaged error (mae) in the cost function $L$, Eq. (6), for learning. Results are shown for (a) the Pearson correlation $\rho_P$, (b) the susceptibility $\chi_4$ and (c) the four-point structure factor $S_4(q,t)$ at times $t = \tau_\alpha^{\mathrm{BB}}/30$ and $t = \tau_\alpha^{\mathrm{BB}}/3$.

portant to use more complex network structures and loss functions.

## IV. SNAPSHOTS

In Fig. 3a of the main manuscript a few snapshots are shown to visualize the performance of GlassMLP. The snapshots are taken from configurations with 25,800 particles. Here, we show some further snapshots for various temperatures and different color codes.

Figure 3 uses the same color code as the main manuscript to visualize dynamic heterogeneity. The isoconfigurational average of the bond-breaking order parameter becomes significantly more heterogeneous at lower temperature and longer times. In particular, the contrast between active and passive regions increases. An interesting observation is also the clear growth with time of rearranged clusters. For $T = 0.23$ at $t = \tau_\alpha^{\mathrm{BB}}/3$ the geometric features visible at this time can be easily traced back to earlier times, showing that they result from individual small clusters which grow and merge. This observation is strongly connected to similar observations in deeply supercooled 2D polydisperse samples [12].

Another visualisation is offered in Fig. 4 where all particles with smaller-than-average propensity are shown in red, the remaining particle being blue. These snapshots emphasize both the time evolution of the clusters for a given temperature and the evolution of the characteristic shape and geometry across temperatures. An obvious effect is the formation of much more pronounced and clearer boundaries between active and passive regions. This effect leads to the higher-order terms in the four-point structure factor presented in Fig. 4 of the main text. Importantly, this also reveals the relatively weak temperature dependence of the characteristic length scale of these domains at $t = \tau_\alpha^{\mathrm{BB}}/3$.

## V. FOUR-POINT CORRELATION FUNCTION

We complement the analysis performed in the main manuscript on the four-point dynamic structure factor $S_4(q, t)$ with the calculation of the four-point correlation function measured in real space,

$$G_4(r;t) = \frac{A}{N_1}\left\langle \sum_{i,j \in N_1} \delta\mathcal{C}_B^i(t)\delta\mathcal{C}_B^j(t)\delta\left[\mathbf{r} - \mathbf{R}_{ij}\right]\right\rangle, \quad (8)$$

with $\delta\mathcal{C}_B^i(t) = \mathcal{C}_B^i(t) - \langle\mathcal{C}_B(t)\rangle$. By definition, the four-point correlation function decays to 0 in the limit $r \to \infty$, with the functional form $G_4(r;t) \sim \exp\left[r/\xi(t)\right]/\sqrt{r}$ expected in 2D. Using the length scales $\xi$ extracted from $S_4(q,t)$ we find a very good agreement between the long-range decay of $G_4(r;t)$ and the expected exponential decay, see Fig. 5. Even more importantly, there is good agreement between MD and GlassMLP for both system sizes. Notice that the spatial decay of the correlations

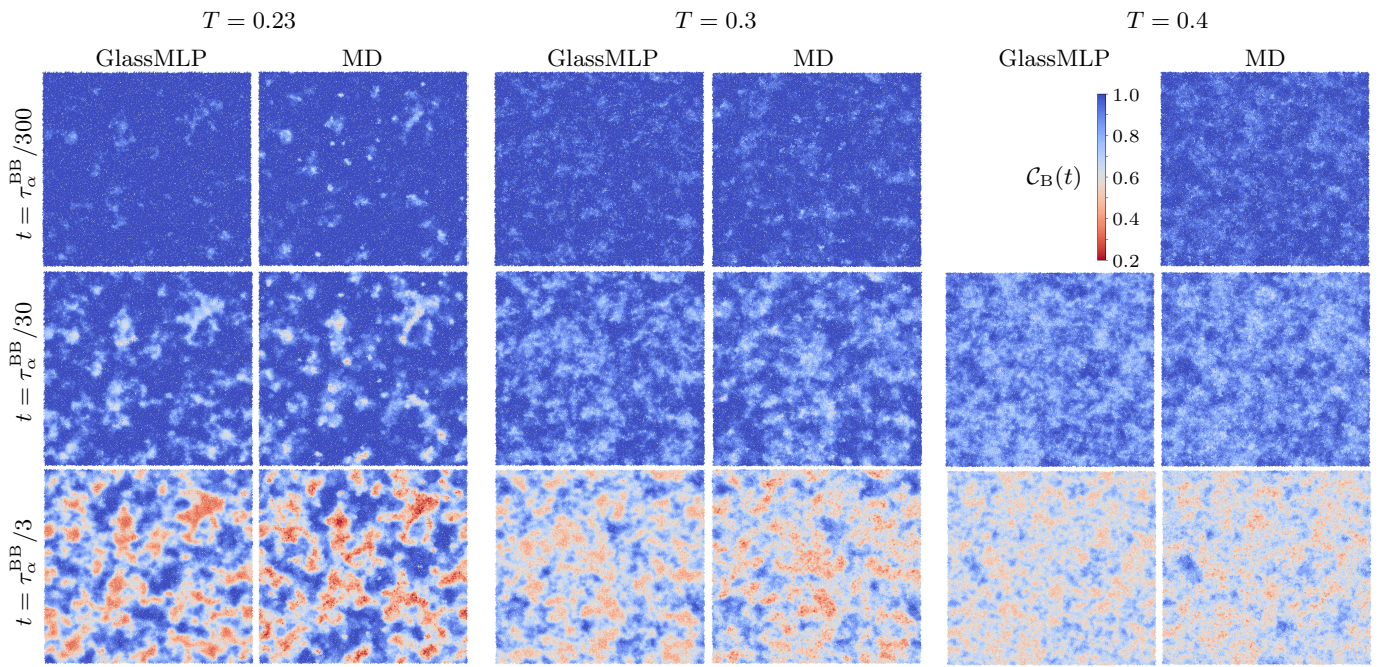

FIG. 3. Snapshots as in Fig. 3a of the main manuscript, shown for different temperatures and a wider range of time scales.

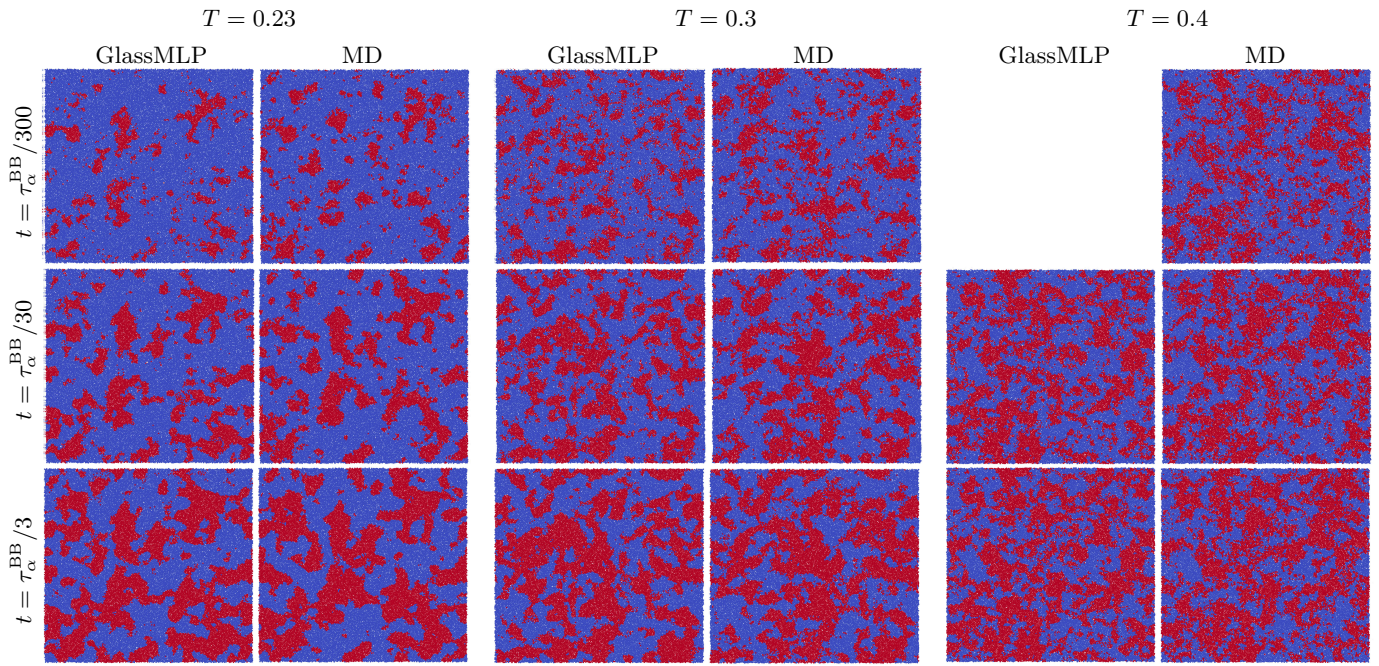

FIG. 4. Same data as in Fig. 3, with different color code. For each configuration, the average propensity $\mathcal{C}_B(t)$ over all particles is calculated. Particles with propensity smaller than this average are shown in red, all other particles are blue.

evolves weakly with temperature, but the absolute amplitude increases by an order of magnitude towards low temperature over the studied range, following the trend seen for the dynamic susceptibility $\chi_4$.

## VI. FINITE SIZE ANALYSIS

In the main manuscript we have used the scalability in system size of the trained GlassMLP to determined the dynamical correlation length $\xi$ from the four-

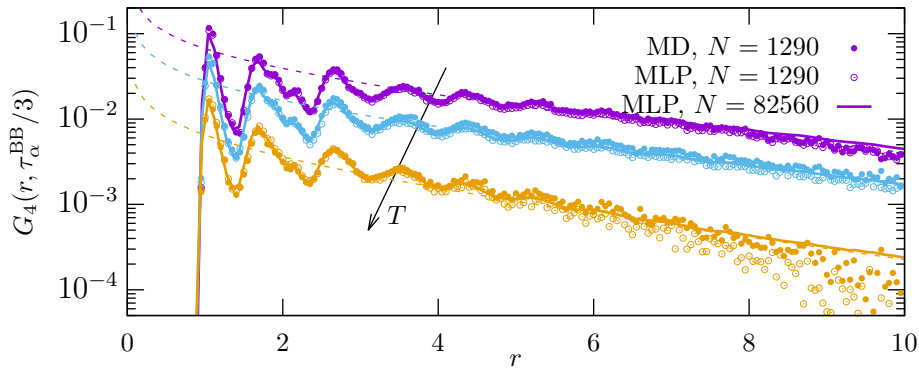

FIG. 5. Same as Fig. 4(a) of the main manuscript but for the four-point correlation function $G_4(r, \tau_\alpha^{\rm BB}/3)$. Dashed line shows long-range decay $\sim \exp[r/\xi(t)]/\sqrt{r}$ with $\xi(t)$ as extracted in the main manuscript from $S_4(q,t)$. The dashed and full line nearly perfectly overlap for $r > 4$.

point dynamical structure factor $S_4(q,t)$ extract in large systems. A rigorous finite size analysis presented in Refs. [13, 14] provides extensive evidence that the bond-breaking correlation is indeed not affected by the finite system size, if the system is larger than 1000 particles in two-dimensional systems. In Fig. 6 we provide further evidence for the transferability of the trained GlassMLP networks in a temperature regime which we can still access via molecular dynamics simulations. The results clearly indicate that no systematic dependence of the results on the system size can be observed. For larger systems, it is notoriously difficult to calculate precise values for the susceptibility $\chi_4$ but within the statistical uncertainty the results are consistent. In fact it can be observed that the predictions of GlassMLP are slaved to the MD results and follow the same trends. This highlights that GlassMLP is indeed able to detect dynamic heterogeneity solely from the structural properties of the samples.

The four-point susceptibility $S_4(q, \tau_\alpha^{\rm BB}/3)$ also clearly does not show any systematic finite size effects. There is only a small shift between the GlassMLP predictions and the MD results originating from the underestimation of the susceptibility $\chi_4$. Overall we can thus conclude that the results above mode-coupling temperature $T_{\rm MCT}$ do not show any dependence on the system size and transferability of the trained GlassMLP networks is possible. Since the growth of the dynamic heterogeneities between $T = 0.3$ and $T = 0.23$ is limited and thus likely the situation does not change drastically between these two temperatures, we believe to have sufficient evidence to support the findings in the main manuscript.

## VII. COMPUTER SIMULATIONS: 3D KOB-ANDERSEN MIXTURE

The 3D Kob-Andersen mixture studied in the last part of the manuscript is a non-additive mixture of two types with $\alpha, \beta = \{1, 2\}$ and $\epsilon_{11} = 1.0$, $\epsilon_{12} = 1.5$, $\epsilon_{22} = 0.5$ and $\sigma_{11} = 1.0$, $\sigma_{12} = 0.8$, $\sigma_{22} = 0.88$. The potential is the same as defined in Eq. (1). We further use $C_0 = C_2 = C_4 = 0$ and $r_{\alpha\beta}^{\rm cut} = 2.5\sigma_{11}$. This system is the same as simulated in Ref. [11] and we use the simulation data provided by the authors of this reference.

From the absolute displacements of each particle, $\Delta_i(t) = |\boldsymbol{R}_i(t) - \boldsymbol{R}_i(0)|$ we calculate the isoconfigurational average $\mathcal{R}^i(t) = \langle \Delta_i(t) \rangle_{\rm iso}$ over $M_R = 30$ different replicas. The learning of GlassMLP is then performed identically to the procedure described in the main manuscript. To extract the predictions of the graph neural network (GNN) proposed in Ref. [11] we use their uploaded learned models. For the original MD results, as well as the predictions of the two neural networks, the susceptibility $\chi_4(t) = N_1^{-1}\left(\langle \bar{C}_\mathcal{R}^2(t)\rangle - \langle \bar{C}_\mathcal{R}(t)\rangle^2\right)$ is then calculated from the overlap function $\bar{C}_\mathcal{R}(t) = \sum_{i \in N_1} \tanh\left(20(\mathcal{R}^i(t) - 0.44) + 1\right)/2$. Despite the slightly different functional form, the results are basically identical to the overlap used in Ref. [11]. We have chosen this differentiable form such that we can insert it into the definition of the correlation function (6), required for the loss function discussed in Sec. II B.

## VIII. PROPENSITY DISTRIBUTION AND KULLBACK-LEIBLER DIVERGENCE

In Fig. 2b of the main manuscript we have shown the distributions of the predicted and simulated propensities and found good agreement between both quantities. Here, we quantify this agreement and calculate the Kullback-Leibler (KL) divergence [15],

$$KL(p_{\rm MLP}(\mathcal{X}), q_{\rm MD}(\mathcal{X})) = \int_0^1 d\mathcal{X}\, p_{\rm MLP}(\mathcal{X}) \log \frac{p_{\rm MLP}(\mathcal{X})}{q_{\rm ML}(\mathcal{X})}. \quad (9)$$

Results for the KA2D model are shown in Fig. 7a in which the results for GlassMLP are also compared to Ridge re-

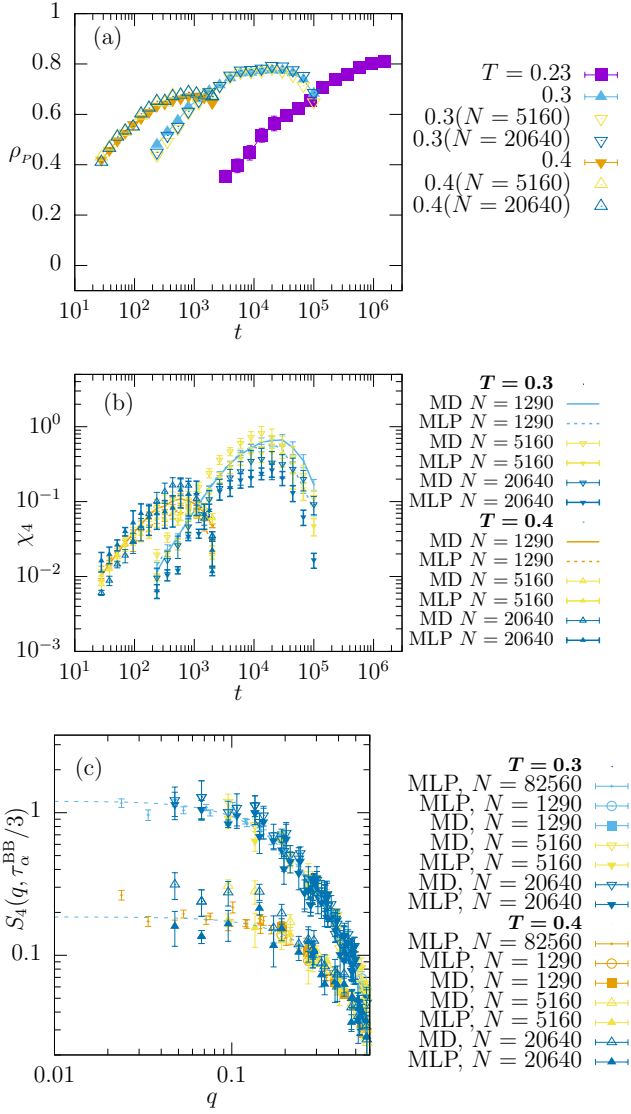

FIG. 6. Finite size analysis of the results presented in the main manuscript for $T = 0.3$ and $T = 0.4$. (a) Pearson correlation coefficient $\rho_P$ as in Fig. 2(a) of the main manuscript. In addition, the trained GlassMLP at $N = 1290$ is applied to larger systems of $N = 5160$ and $N = 20640$ particles. (b) Four-point susceptibility $\chi_4$ as in Fig. 3(b) of the main manuscript. The figure also shows results from molecular dynamics simulations and the transferred GlassMLP results. (c) Four-point structure factor $S_4(q, \tau_\alpha^{\rm BB}/3)$ as in Fig. 4(a) of the main manuscript.

distribution of propensities in the 3D KA mixture of Ref. [11]. Also for this system we observe that the KL divergence between the distributions predicted by GlassMLP and the MD results is well below 0.05 at the structural relaxation time. For short times, the KL divergence is much larger due to the short-time displacements which are not well predicted by GlassMLP. (Note that we use a different definition for propensities in the KA model, as discussed in Sec. VII above.) Similar to the observations discussed in Fig. 5 in the main manuscript for the Pearson correlation coefficient we find that the GNN performs nearly perfectly in the short-time limit while it does significantly worse than GlassMLP at the structural relaxation time [11].

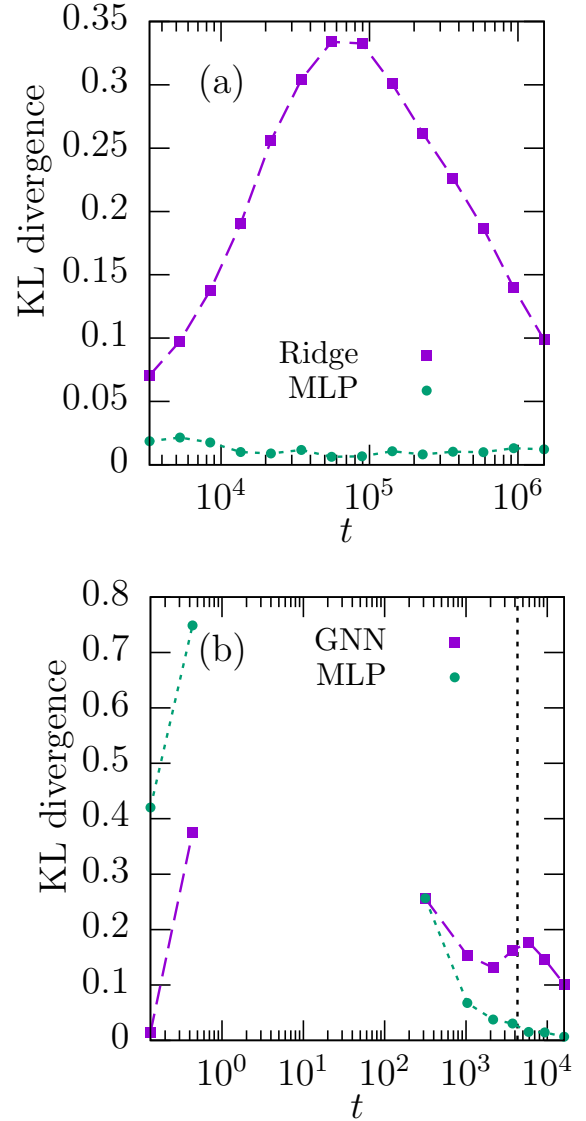

FIG. 7. Kullback-Leibler divergence, as defined in Eq. (9), between predicted propensity distributions and the MD simulations for, (a) the KA2D model at $T = 0.23$, and (b) the KA model.

gression. The KL divergence clearly confirms the optical impression and highlights that GlassMLP very well predicts the underlying distribution of propensities. Importantly, while an excellent Pearson correlation coefficient could also be achieved using linear Ridge regression, the KL divergence shows that the probability distributions are not very well described by this technique.

In Fig. 7b we additionally provide quantitative comparisons between the KL divergence extracted for the



\end{document}